\documentclass[a4paper,11pt]{article}
\pdfoutput=1
\usepackage{jcappub}
\usepackage{graphicx}
\usepackage{caption}
\usepackage{subcaption}
\usepackage{placeins}

\newcommand{\er}[1]{(\ref{#1})}          
\newcommand{\mpl}{M_{\rm Pl}}

\newcommand{\pps}{\varphi'^2}

 \newcommand{\showlabel}[1]{
   \label{#1}
 }

\title{How to Avoid a Swift Kick in the Chameleons}

\author[a]{Antonio Padilla,}
\author[b]{Emma Platts,}
\author[a]{David Stefanyszyn,}
\author[b]{Anthony Walters,}
\author[b]{Amanda Weltman,}
\author[a]{Toby Wilson}

\affiliation[a]{School of Physics and Astronomy, University of Nottingham, Nottingham NG7 2RD, UK}
\affiliation[b]{Astrophysics, Cosmology and Gravity Centre, Department of Mathematics and Applied Mathematics, University of Cape Town Rondebosch 7701, Cape Town, South Africa}

\emailAdd{antonio.padilla@nottingham.ac.uk}
\emailAdd{pltemm002@myuct.ac.za}
\emailAdd{ppxds1@nottingham.ac.uk}
\emailAdd{tony.walters@uct.ac.za}
\emailAdd{amanda.weltman@uct.ac.za}
\emailAdd{toby.wilson@nottingham.ac.uk}
\abstract{Recently, it was argued that the conformal coupling of the chameleon to matter fields created an issue for early universe cosmology. As standard model degrees of freedom become non-relativistic in the early universe, the chameleon is attracted towards a ``surfing" solution, so that it arrives at the potential minimum with too large a velocity. This leads to rapid variations in the chameleon's mass and excitation of high energy modes, casting doubts on the classical treatment at Big Bang Nucleosynthesis. Here we present the DBI chameleon,  a consistent high energy modification of the chameleon theory that dynamically renders it weakly coupled to matter during the early universe thereby eliminating the adverse effects of the `kicks'.   This is done without any fine tuning of the coupling between the chameleon and matter fields, and retains its screening ability in the solar system. We demonstrate this explicitly with a combination of analytic and numerical results.}

\begin{document}
\maketitle

\section{Introduction}
\label{intro}
Many models of cosmology involve light scalar degrees of freedom. These scalars often couple to the trace of the energy momentum tensor with gravitational strength and can therefore mediate long range forces comparable to gravity. Their existence in Nature can be argued for from two view points. The first is a bottom-up approach, where they are introduced to solve outstanding problems in cosmology such as the origin of the late time cosmic acceleration (for several examples, see \cite{DEreview,MGreview}). In this case they are assumed to exist in the low energy effective field theory (EFT) of an unknown UV completion of gravity. They can also be motivated from a top-down approach, for example as light moduli coupled to gravity in string compactifications (see e.g. \cite{conlon} for a discussion of  moduli in string theory).  Of course, if these fields are to play an interesting role for cosmology, their masses must be of order the Hubble scale or smaller in order to tame any Yukawa suppressions. The scalar moduli in string theory are typically exactly massless.  Scalars this light  can  create problems for local phenomenology and with the tight constraints coming from Big Bang Nucleosynthesis (BBN). It is well known that Einstein's equations are sufficient to describe solar system gravitational physics so the addition of a canonical scalar field that couples to matter with gravitational strength will provide an $\mathcal{O}(1)$ deviation from  General Relativity that would be ruled out experimentally \cite{Will}.  If light scalar fields are to play an interesting role in the gravitational interaction at cosmological scales, a mechanism is required to make them inert locally. 

The chameleon theory \cite{cham1,cham2} was originally conceived to solve the moduli stability problem in string theory. A conformal coupling to matter ensures that the potential experienced by a chameleon is not just $V(\phi)$ but also depends on the matter density of the environment. It follows that the mass of the chameleon becomes light in regions of low density and heavy in high density environments. A large mass reduces the Compton wavelength of the field and makes the force mediated by exchanges of chameleons short range. Even though there is no known chameleon model to explain dark energy \cite{Kaloper:2007gq,chamnot}, it provides an interesting mechanism to shut down long range scalars in the solar system and therefore warrants further study. The simplest chameleon model with all matter fields coupled to gravity in the same way, is given by
\begin{equation}
S = \int d^{4}x \sqrt{-g} \left[ \frac{M_{\rm Pl}^{2}}{2}R - \frac{1}{2} (\partial \phi)^{2} - V(\phi)\right] + S_{m}\left[\tilde{g}_{\mu\nu}, \psi\right],  \label{chamaction}
\end{equation}
where $\tilde{g}_{\mu\nu} = \exp(2\beta \phi / M_{\rm Pl})g_{\mu\nu}$ and $\beta$ is a dimensionless coupling.  When $\beta \sim \mathcal{O}(1)$ the chameleon couples to matter with gravitational strength. Provided the bare potential and the conformal factor slope in opposite directions, this introduces an environmental minimum at $\phi_\text{min}$. Once the  the chameleon settles there, it develops an effective mass $m^2_\text{eff}(\phi_\text{min}) = V_\text{eff}''(\phi_\text{min})$ that generically increases with the local value of $\rho-3p$, where $\rho$ and $p$ are the background Einstein frame energy density and pressure.

Naively, in a radiation dominated epoch the conformal coupling to matter would ensure that any interactions between the chameleon and matter fields in the Jordan frame would be indirect and only possible through mediation of virtual gravitons.  However, in a more realistic model the quantity $\Sigma = (\rho - 3p)/ \rho$ can temporarily become non-zero in a radiation dominated universe whenever a massive particle species becomes non-relativistic. This generates a number of \textit{kicks} in $\Sigma$ as a function of {\it Jordan frame} temperature when one accounts for all the Standard Model (SM) particles \cite{cham3, kick1}. The deviation away from zero can be of $\mathcal{O}(0.1)$ and therefore have significant impact on the scalar's evolution. These peaks in the $\Sigma$ spectrum occur when the temperature of the radiation bath is approximately equal to the mass of a given particle species. Prior to the kick, when the temperature is larger than the mass, the particle is relativistic and does not contribute to $\Sigma$. The contribution after the peak is also small. In this case the temperature is less than the mass of the particle and $\Sigma$ decreases due to Boltzman suppression.  

The consequences of these kicks were first analysed in \cite{cham3} and later in \cite{kick1,kick2} where quantum effects are included. It was shown that for the theory described by (\ref{chamaction}), the kicks have an adverse effect on the chameleon cosmology. In particular, they found a \textit{surfer solution}, characterised by constant Jordan frame temperature, which generically drives the chameleon to the minimum of its effective potential where it arrives with a very high velocity, dictated by the radiation temperature. The field then climbs the steep side of its effective potential until the reflection point, where it has zero kinetic energy, and rolls back down. This process generates rapid variations in the chameleon's effective mass and the short time scale of the process excites highly energetic modes. Dissipative effects then invalidate any classical treatment of the chameleon's evolution with the energy stored in fluctuations becoming $\mathcal{O}(1)$ in units of the energy stored in the background field.  According to \cite{kick1,kick2}, this may indicate the break down of the EFT describing the chameleon's evolution and the model may then fail to be a classical field theory at BBN. As long as $\beta > 1.82$, the surfing solution will exist for generic initial conditions. For $0.42<\beta<1.82$, there is no surfer,  but the chameleon will still arrive at the potential too quickly, at least if we assume a BBN bound on the field value prior to the last kick. It would appear that the only way to avoid these issues is to tune $\beta$ such that the chameleon is sufficiently weakly coupled to matter fields.

Of course,  we could take the view that the chameleon EFT is not valid as far back as BBN, in which case the analysis of  \cite{kick1,kick2}  does not apply. Indeed,  in \cite{justin} it has been suggested that  a low cutoff for the EFT may even be desirable when one takes into account corrections to the chameleon mass from matter loops, going like $\Delta m^2 \sim \beta^2 \mu^4/M_{\rm Pl}^2$, where $\mu$ is the cutoff.  However,  all such corrections to a vanishing potential can be associated with radiative corrections to the cosmological constant in Jordan frame\footnote{We thank N. Kaloper for pointing this out.}. This is due to the fact that,  in the limit of vanishing potential, the scalar only mixes {\it kinetically} with the Jordan frame graviton. Similar considerations guarantee that there is no violation of the weak equivalence principle induced by matter loops \cite{fujii,nicolis,piazza}.  In any event, we take the view that any credible alternative to General Relativity should remain valid as far back as BBN, such that it remains predictive in the testable regime.

The conclusions of \cite{kick1,kick2} require us to place further constraints on the strength of the chameleon-matter coupling. If such constraints were to hold they would challenge the relevance of lab based tests of the chameleon \cite{clare,berk,neutron,cannex}.  The purpose of this paper is to demonstrate that a consistent UV modification of the chameleon action (\ref{chamaction}) can  stabilise the scalar's evolution in the presence of SM kicks, removing the need to strongly constrain the matter coupling.  Indeed,  at high energies, the UV correction helps to {\it dynamically} renormalise the coupling strength  rendering the chameleon weakly coupled to matter fields during BBN (and therefore suppressing the surfing behaviour). At low energies, in the late universe, we recover the original chameleon set up (\ref{chamaction}), with no need to strongly constrain $\beta$ from early universe considerations.

 The UV correction we will consider is in the form of derivative self interactions such that the kinetic term describing scalar fluctuations picks up a large $Z$ factor on non-trivial scalar backgrounds, $-\frac12 Z ( \partial \delta \phi)^2$. It follows that when the scalar fluctuation is canonically normalised, it couples to the trace of the energy momentum tensor with strength as  $\beta /\sqrt{Z} M_{\rm Pl}$. In the regime where the derivative interactions dominate over the canonical kinetic term, the $Z$ factor becomes large and the coupling to matter becomes weak. This is essentially how the Vainshtein mechanism works \cite{vain1,vain2,vain3,vainunit}. Just like the chameleon mechanism, the Vainshtein effect renders the interactions between the scalar and matter sources negligible without tuning $\beta$. Usually, the Vainshtein mechanism is employed on static backgrounds where large $Z$ factors suppress scalar forces in the solar system. However, here we are considering cosmology where homogeneous backgrounds are relevant which is an important distinction in terms of being able to construct a consistent EFT. In particular, it allows us to avoid the well known pathologies associated with screening on static backgrounds. There, the coefficients of leading irrelevant operators must violate the positivity condition required for a Lorentz invariant UV completion \cite{positivity} if the non-linearities are to consistently dominate the dynamics. This has been linked to the appearance of superluminal fluctuations in the low energy theory. The opposite is true for homogeneous configurations \cite{vainunit}. The condition for consistent non-linear solutions now coincides with the positivity condition, ensuring that the interactions could come from a Lorentz invariant UV theory and can therefore form a sensible effective description of an unknown high energy completion. It is important that any corrections we add to (\ref{chamaction}) do indeed represent a healthy EFT because one of the major issues we are trying to address is the break down of calculability in the original model.

A generic correction to (\ref{chamaction}) of the form
\begin{equation}
\delta \mathcal{L} \sim c_{n}\frac{(\partial \phi)^{2n}}{\Lambda^{4n-4}} \label{correction}
\end{equation}
would be problematic. As soon as the first irrelevant operator became important an infinite tower of operators would too, with the dynamics dominated by the $n \rightarrow \infty$ limit. Each would come with arbitrary coefficients and calculability would be lost as soon as the $Z$ factor becomes large. We therefore choose a correction where the non-linearities can be kept under control within an effective description. The theory we study is  governed by a  Dirac-Born-Infeld (DBI) generalisation of the chameleon, 
\begin{equation}
S = \int d^{4}x \sqrt{-g} \left[ \frac{M_{\rm Pl}^{2}}{2}R + \Lambda^{4} - \Lambda^{4}\sqrt{1+\frac{(\partial \phi)^{2}}{\Lambda^{4}}} - V(\phi)\right] + S_{m}[\tilde{g}_{\mu\nu}, \psi], \label{DBI}
\end{equation}
where $\tilde{g}_{\mu\nu}$ is as defined previously. Expanding the action to leading order in the low energy limit $(\partial \phi)^{2} \ll \Lambda^{4}$ recovers the original chameleon model such that at low energies the main features of the chameleon are retained and the derivative interactions are not required to suppress forces in the solar system. It is easy to see at next order that the leading irrelevant operator has a positive coefficient,  ensuring that the $Z$ factor begins to grow  on non-trivial homogeneous backgrounds,  as desired \footnote{If we choose the `wrong sign' for the leading irrelevant operator then the Z factor grows on static backgrounds and the theory can be used to screen scalar forces in the solar system \cite{dbionic}}.  Since it is absolutely crucial for us to retain calculational control of our theory, we emphasize that the structure of the DBI kinetic term is protected by symmetry, so its effective description can be trusted even when non-lineariries become important.  The DBI structure comes with an added bonus, imposing a speed limit on the chameleon that helps to suppress dangerously large field excursions. Again, we emphasize that although a generic K-essence theory would also generate the desired large $Z$ factors, it is only the DBI structure that does so whilst remaining a trustworthy EFT.

The rest of this paper is organised as follows: we begin with a recap of DBI theory in section \ref{sec:DBI} before reproducing the results of \cite{kick1,kick2} in Section \ref{chameleonsec}, using a dynamical systems approach to show the existence of the troublesome surfing solution in the chameleon limit. We shall also  solve the equations of motion numerically  in the presence of a kick, explicitly demonstrating the inevitability of the surf at too strong a coupling. In Section \ref{DBIsec} we show that the DBI-chameleon nullifies the surfer solution for sufficiently small values of $\Lambda$ relative to the energy scale of the kick. Again, our conclusions are derived from a dynamical systems description and from numerically solving the equations of motion in the presence of a non constant $\Sigma$.  We also explicitly demonstrate how the DBI scale can be used to suppress field excursions during the kicks. These effects are easily understood in terms of a weakening of the effective coupling to matter at high energies, entirely analogous to Vainshtein screening,  along with the speed limit imposed by the DBI structure.  We conclude in Section \ref{conclusionsec},  having demonstrated that  the DBI correction weakens the effective coupling to matter at high energies in a dynamical but controllable way, releasing the chameleon from the constraints on the matter coupling derived in \cite{kick1,kick2}. 

\section{DBI Recap and Quantum Corrections} \label{sec:DBI}
Let us motivate the action in (\ref{DBI}) further and make its healthy properties manifest. Taking the limit of decoupled gravity ($M_{\rm Pl} \rightarrow \infty$) and dropping $V(\phi)$, the theory is built from an infinite tower of $(\partial \phi)^{2n}$ operators and describes an interacting scalar propagating on a Minkoswki background. It has the same form as (\ref{correction}) but, unlike many theories involving the Vainshtein mechanism,  the coefficient of each operator is set by an additional  symmetry,  specifically
\begin{equation} \label{sym}
t\to \frac{1}{\sqrt{1-v^2}} \left(t-v \frac{\phi}{\Lambda^2}\right), \qquad \phi \to\frac{1}{\sqrt{1-v^2}} (\phi- \Lambda^2  vt).
\end{equation}
 Formally, this theory describes the scalar sector of the Dirac-Born-Infeld action, whose uniqueness with respect to the symmetry ensures that its structure is not spoilt by quantum corrections. As long as the proper acceleration of $\phi$ is kept small in appropriate units, it describes a healthy EFT even in the limit $(\partial \phi)^{2} \rightarrow \Lambda^{4}$, unlike a generic K-essence model (see \cite{Riding} for an alternative view of generic K-essence theories, relying on a partial ERG analysis). Indeed, it is the only K-essence model with an enhanced soft limit \cite{soft}. 
 
 The presence of irrelevant operators ensures that the theory possess a non-trivial $Z$ factor. Expanding to quadratic order about a homogeneous background solution $\phi(t)$ we find $\delta \mathcal{L} \sim \frac12 Z^{\mu\nu} \partial_\mu {\delta \phi} \partial_\nu {\delta \phi}$ where 
\begin{equation}
Z^{tt} = \left(1 - \frac{\dot{\phi}^{2}}{\Lambda^{4}}\right)^{-3/2}, \qquad Z^{ti}=0, \qquad Z^{ij}=\left(1 - \frac{\dot{\phi}^{2}}{\Lambda^{4}}\right)^{-1/2} \delta^{ij},
\end{equation}
and $\delta \phi$ is a small fluctuation. Clearly $Z^{\mu\nu}$ gets large for $\dot{\phi} \rightarrow \Lambda^{2}$ at which point the fluctuation couples weakly to matter, as desired. The theory becomes strongly coupled at $\Lambda$ where we expect new heavy physics to take over. 

The symmetry that sets each coefficient can be understood from a higher dimensional picture \cite{reunited}. Consider a probe brane localised at $y = \phi(x^{\mu})/\Lambda^2$ in a Minkowski bulk where $y$ is the coordinate of the fifth dimension. The effective description of the brane depends on the induced metric, $\hat{g}_{\mu\nu} = \eta_{\mu\nu} + 
\frac{\partial_{\mu}\phi \partial_{\nu}\phi}{\Lambda^4}$, and the leading order contribution to the brane action is
\begin{equation}
\label{brane}
S =-\Lambda^4 \int d^{4}x \sqrt{\hat{g}} =- \Lambda^4 \int d^4 x \sqrt{1+ \frac{(\partial \phi)^{2}}{\Lambda^4}}
\end{equation}
where we identify $\Lambda^4$ with the brane tension. This is symmetric under five dimensional  Lorentz transformations and is the origin of the non-linearly realised symmetry (\ref{sym}).

Of course, (\ref{brane}) is a truncation of the brane action and corrections to it will also generate $\phi$ operators invariant under the non-linear symmetry (\ref{sym}). The corrections will involve derivatives of the induced metric, and therefore second derivatives on $\phi$, and are encoded in increasing powers of the brane's extrinsic curvature $K^{\mu}_{\nu}$, with the index raised using $\hat g_{\mu\nu}$. 
Any Lorentz invariant operator built out of $K^{\mu}_{\nu}$ is guaranteed to be symmetric and could, in principle, be generated by quantum corrections. For example, at first order the only Lorentz invariant contraction is simply the trace $K$ and at quadratic order we can generate $K^\mu_\nu K^\nu_\mu$ and $K^{2}$. The argument extends trivially to higher order contractions and of course involves more terms at every order. The truncation relevant for this paper (\ref{brane}) is zeroth order in the extrinsic curvature and to be a consistent limit we must ensure that all higher order operators are suppressed, in other words $K^{\mu}_{\nu} \ll \Lambda$ where
\begin{equation}
K^{\mu}_{\nu} = \frac{\gamma}{\Lambda^2} \nabla^{\mu}\nabla_{\nu}\phi - \frac{\gamma^{3}}{\Lambda^{6}} \nabla^{\sigma}\phi  \nabla^{\mu}\phi  \nabla_{\nu}\nabla_{\sigma} \phi 
\end{equation}
and $\gamma=1/\sqrt{1+ \frac{(\partial \phi)^{2}}{\Lambda^4}}$. On the cosmological background of interest in this paper we must therefore satisfy the following conditions
\begin{equation}
\frac{\gamma H \dot{\phi}}{\Lambda^{3}} \ll 1, \qquad \frac{\gamma^{3} \ddot{\phi}}{\Lambda^{3}} \ll 1
\label{higherordercorrections}
\end{equation}
throughout evolution such that the truncation to the DBI square root is consistent. As we will see later on, this will indeed be case, even in a regime where the DBI term generates a large Z factor, and the scalar fluctuations become weakly coupled to matter.  Note that the DBI structure starts to be felt when, $|\dot \phi| \sim \Lambda^2$, whilst initially maintaining $\gamma \sim \mathcal{O}(1)$. Assuming that $\ddot \phi \sim H \dot \phi$, in the DBI  regime the conditions reduce to $H \ll \Lambda$.

In the full action (\ref{DBI}) we are considering, the symmetry (\ref{sym}) is broken by the potential $V(\phi)$ and at finite $M_{\rm Pl}$ and one could worry that the structure of the DBI kinetic interactions is spoilt by loops. Quantum corrections arising from the latter will necessarily be Planck suppressed, and can be consistently neglected provided  $\Lambda \ll M_{\rm Pl}$. The importance of quantum corrections associated with $V(\phi)$ depends on the strength of those interactions. If we assume that the potential decays at large values of $\phi$ (as is often done when evoking the chameleon mechanism) then the mass scale $M$ controlling the strength of these interactions enters the action with positive powers. These operators then only break the symmetry weakly as long as $M \ll \Lambda$ and when this condition holds, symmetry breaking operators associated with $V(\phi)$ can also be consistently ignored. We note that this condition is naturally satisfied since the size of $\Lambda$ is controlled by early universe physics and $M$ by late universe physics. The DBI structure is therefore stable against loops as long as $M \ll \Lambda \ll M_{pl}$.

\section{Surfing Chameleon} \label{chameleonsec}
We shall now derive the chameleon surfer using a dynamical systems analysis, which complements the original approach in \cite{kick1,kick2}.  In terms of our generalised set-up this corresponds to the limit of decoupling the DBI derivative interactions by taking $\Lambda \rightarrow \infty$. For a flat FLRW cosmology, the chameleon model (\ref{chamaction})  yields the following equations of motion in the Einstein frame, 
\begin{equation}
3M_{\rm Pl}^2H^2 = \rho_\phi + \rho, 
\label{elimination}
\end{equation}
\begin{equation}
M_{\rm Pl}^2\left(2\dot H+3H^2\right) = -p_\phi -p,
\end{equation}
\begin{equation}
\ddot{\phi} + 3H\dot{\phi} + V'(\phi) = -\dfrac{\beta}{M_{\rm Pl}}\rho \Sigma,
\end{equation}
where $H=\dot a/a$ is the Hubble parameter, $\rho$ and $p$ the energy density and pressure of matter, and  $\rho_\phi=\frac{\dot \phi^2}{2} +V, ~p_\phi=\frac{\dot \phi^2}{2} -V$, the energy density and pressure  associated with  the scalar field. As in \cite{kick1,kick2}, we denote $\Sigma = (\rho - 3p)/ \rho$, so that for a perfect fluid with equation of state $w=p/\rho$, we have $\Sigma = 1-3w$.  Also in line with the analysis of \cite{kick1,kick2}, we shall henceforth neglect the potential $V(\phi)$ by assuming that in the early universe the matter fields are the dominant source of the chameleon's dynamics. This is, of course, a reasonable assumption, as kicks become truly relevant in the case of an `undershoot' solution, which inherently has $\rho$ dominate over $V$ \cite{cham3}. This approximation will begin to break down as the chameleon approaches the minimum of its effective potential as dictated by the surfing solution. Using (\ref{elimination}) to eliminate $\rho$ from the equations, we are left with 
\begin{equation}
M_{\rm Pl}^2H^2(4-\Sigma) + 2M_{\rm Pl}^2\dot{H} + \frac{(2+\Sigma)}{3}\rho_\phi = 0, \showlabel{basicH}
\end{equation}
\begin{equation}
\ddot{\phi} + 3H\dot{\phi} + 3M_{\rm Pl}\beta\Sigma H^2\left(1-\dfrac{\rho_\phi}{3M_{\rm Pl}^2H^2}\right) = 0, \showlabel{basicPhi}
\end{equation}
where we have used the fact that $p_\phi=\rho_\phi=\frac{\dot \phi^2}{2}$ when the potential is neglected.
It is at this point we can begin to look for the surfer as a fixed line in an autonomous dynamical system with the following set of variables,  
\begin{eqnarray}
x &=& \beta\phi + M_{\rm Pl}\ln a,
\label{xdef}
\\\
y &=& \dfrac{\rho_\phi}{3M_{\rm Pl}^2H^2},
\label{ydef}
\\
z &=& \beta\dot{\phi} + M_{\rm Pl}H.
\label{zdef}
\end{eqnarray}
We have defined $x$ such that $\dot{x}=0$ represents a constant Jordan frame temperature $T_J \propto e^{-{\beta \phi}/{{M_{\rm Pl}}}}/{a}$, as this represents the characteristic property of the surfer \cite{kick1,kick2}. $y$ is a familiar choice, and is a standard one for fixed point analysis in cosmology, the ratio between the energy density of $\phi$ and the critical density.  The autonomous system of equations is now given by,
\begin{eqnarray}
\dot x &=& z \label{xdot},\\
\dot y &=&  \frac{1}{H_y} \left[H_z \left(3Hz-\frac12 M_{\rm Pl} H^2 (1-y)\left( 2+\Sigma(1-6\beta^2) \right)\right) \right. \nonumber \\
&& \left.\qquad +\frac{H^2}{2} \left(\Sigma-4-y(\Sigma+2)\right) \right], \label{ydot} ~~\\
\dot z &=& -3Hz+\frac12 M_{\rm Pl} H^2 (1-y)\left( 2+\Sigma(1-6\beta^2) \right), \label{zdot}
\end{eqnarray}
where $H_y=\frac{\partial H}{\partial y}$, $H_z=\frac{\partial H}{\partial z}$ and $H(y, z)$ is given implicitly by the equation
\begin{equation}
M_{\rm Pl}^2H^2y =\frac{ \left(z - M_{\rm Pl}H\right)^2}{6\beta^2}.
\label{constraint1}
\end{equation}
In this case we can solve for $H(y,z)$ explictly, 
\begin{equation}
H(y,z) = \dfrac{z}{M_{\rm Pl}(1\pm \sqrt{6\beta^2y})}.
\label{HforCham}
\end{equation}
If we assume that the conformal factor in the matter coupling decreases over time, then  $\frac{\beta \dot \phi}{M_{\rm Pl}}=\frac{z}{M_{\rm Pl}}-H<0$, and so we take the lower root in (\ref{HforCham}).  For $y \neq 1/6 \beta^2$, the fixed points of this system have $z=0$, with $x$ and $y$  arbitrary constants. Such fixed points always have $H=0$, and correspond to an empty Universe.  Of greater relevance to the SM kicks in the early Universe are the additional fixed points  that appear when $\Sigma= 2/(6\beta^2-1)$.  These have  $y =1/6 \beta^2$, $z=0$, with $x$ an arbitrary constant, and, crucially, they place no constraint on $H$. This latter property will be relevant when we generalise our analysis to the full DBI-chameleon model discussed in the introduction.

The line of additional fixed points at the critical value of  $\Sigma$ are precisely the troublesome surfer found in \cite{kick1,kick2}. They have constant Jordan frame temperature $\dot x=0$ and can exist in a Universe at any scale ($H$ arbitrary) provided $\Sigma$ passes through $2/(6\beta^2-1)$. The alert reader will notice that this critical value differs slightly from the one presented in \cite{kick1,kick2}. Their result agrees with ours to leading order in large $\beta$, and is merely an artefact of their approximation.

\begin{figure}[h!]
\centering
\includegraphics[width=\linewidth]{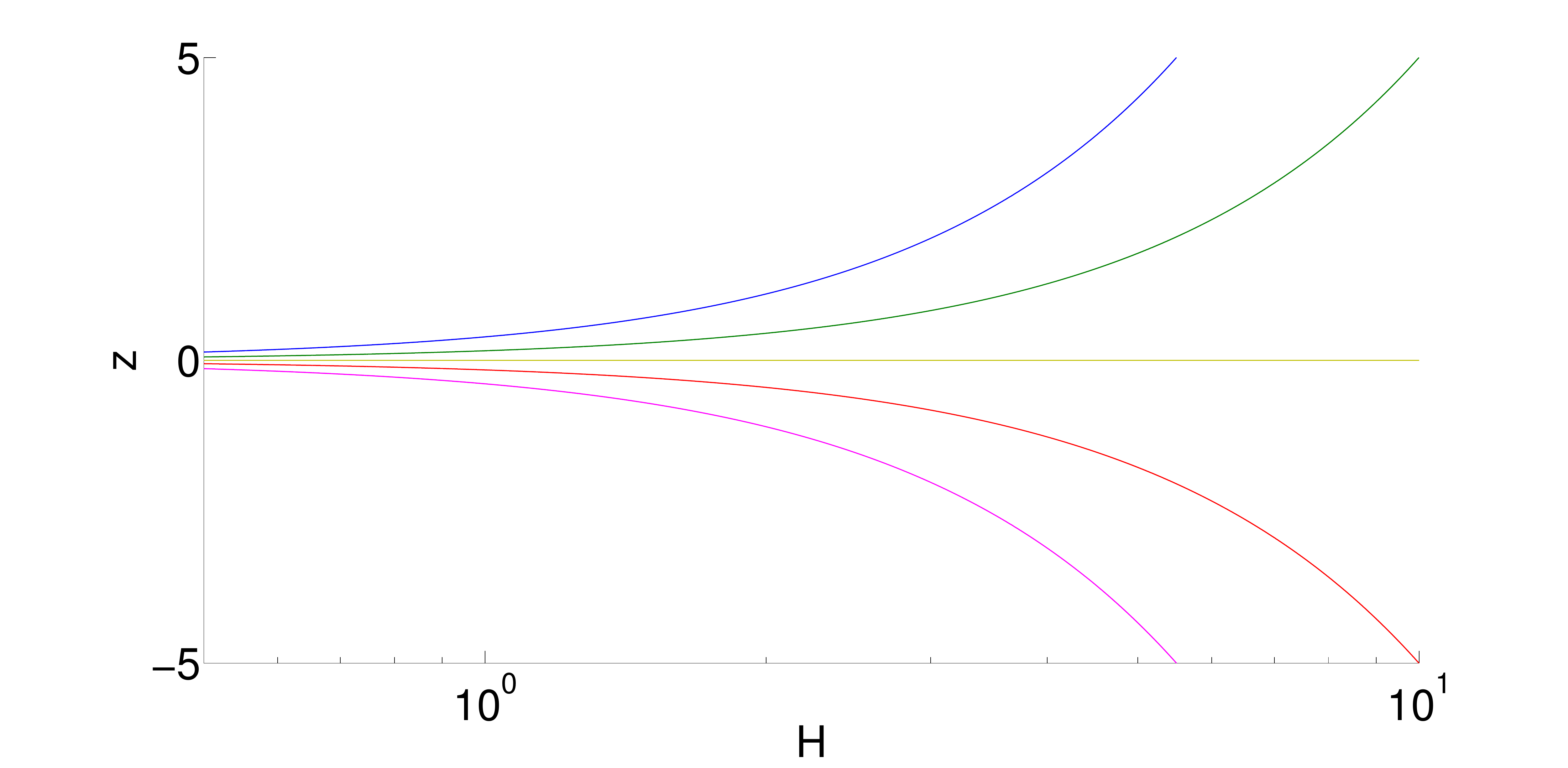}
\caption{Plot of $z$ against $H$ for a range of initial conditions, indicated by different colours. $\Sigma = \dfrac{2}{6\beta^2-1}$. We numerically set $M_p=10000$ and $\beta=1$.}
\label{chamsurf}
\end{figure}

We can solve the equations (\ref{xdot} - \ref{constraint1}) numerically for an array of initial conditions. We plot the trajectories of these solutions in the $(H, z)$ plane in Figure \ref{chamsurf}.
When  $\Sigma = 2/(6\beta^2-1)$ Figure \ref{chamsurf} shows us that for a wide range of initial conditions, the solutions are attracted to the $z=0$ line, even for $H\neq 0$.   We can see this behaviour analytically by expanding the dynamical system about the surfer's trajectory, $x=x_0+\Delta x, ~y=\frac{1}{6\beta^2} +\Delta y, z=\Delta z$.  The equations (\ref{xdot}) to (\ref{zdot}) and the constraint (\ref{constraint1}) yield 
\begin{equation}
\Delta \dot  x \approx \Delta z, \qquad   \Delta\dot y \approx \frac{1}{3M_{\rm Pl} \beta^2} \Delta z, \qquad   \Delta \dot z \approx -3H\Delta z,
\end{equation}
along with $\Delta z=-3 M_{\rm Pl} \beta^2 H\Delta y$.   These are easily solved to give $\Delta x \propto \Delta y \propto \Delta z \propto \exp(-3Ht)$, demonstrating the fact that the solution is indeed attracted to the surfer at late times in an expanding Einstein frame\footnote{There is also  a constant contribution to $\Delta x$ that can be absorbed into a redefintion of $x_0$}.

 If we allow $\Sigma$ to vary from this critical value, we can once again study the dynamical system analytically as an expansion about the surfer, and the critical value of $\Sigma$.  Indeed, in a physical setup, the latter is not a constant, but varies with Jordan frame temperature $T_J =T_{J0} \exp(-x/M_{\rm Pl})$, or equivalently with $x$. This means in a neighbourhood  of the surfer   we can approximate  $\Sigma(x) \approx \frac{2}{6\beta^2-1}+ \Sigma'(x_0) \Delta x$, such that at leading order the dynamical system is reduced to
 \begin{eqnarray}
\Delta \dot  x &\approx& \Delta z, \\ \Delta\dot y &\approx& \frac{1}{3M_{\rm Pl} \beta^2} \Delta z+ H\left(\frac{1}{6\beta^2}-1\right)^2  \Sigma'(x_0) \Delta x, \\  \Delta \dot z &\approx & -3H\Delta z-3 \beta^2 M_p H^2\left(\frac{1}{6\beta^2}-1\right)^2 \Sigma'(x_0) \Delta x.
\end{eqnarray}
The solutions are all of the form ${\cal A}_+ \exp(m_+ t)+{\cal A}_- \exp(m_- t)$, where 
\begin{equation}
m_\pm =-\frac{3H}{2}\left(1 \pm \sqrt{1-\frac43 \beta^2 M_p \left(\frac{1}{6\beta^2}-1\right)^2 \Sigma'(x_0) } \right).
 \end{equation}
 As the Jordan frame temperature cools, the kick function $\Sigma$ will be initially vanishing, before rising to some maximum value and then falling back to zero (see Figure 1 of \cite{kick1}). For certain values of $\beta$, it will pass through the critical value on both the rise and fall.  When it passes through for the final time,  on the fall at some critical temperature $T_J^c(\beta)$,  it does so with $\Sigma'(x_0)=-\frac{T_J^c}{M_{\rm Pl}} \frac{d \Sigma }{dT_J} <0$, ensuring that $m_+$ is real and negative while $m_-$ is real and positive. The latter signals a possible instability for the surfer, but it turns out to be very mild. This is because for physically realistic kicks (see e.g. \cite{kick1,kick2}), we have that $M_{pl} |\Sigma'(x_0)| \sim \Sigma(x_0) =2/(6 \beta^2-1)$ and so $m_- \lesssim H$. Thus the would be instability behaves like a constant during a Hubble time, which again may be reabsorbed into a redefinition of $x_0$. This explains the attractor behaviour demonstrated numerically in \cite{kick1,kick2}.  Here we see the importance of modelling the kicks appropriately. For example,  modelling the kick with a delta function would completely destabilise the would-be surfer on account of the infinite gradient in $\Sigma$. Later on we will model the kick with a simple gaussian whose properties better reflect the physical set-up.
 
We conclude this section by recalling the effect of the surfer on the cosmological evolution of the chameleon.  The existence of this solution and its approximate attractor behaviour allows the chameleon to perform very large field excursions during the critical phase of the kick. This is because the surfer itself corresponds to constant Jordan frame temperature, so once it gets caught in this regime, it stays there since the cooling stops and the kick stays close to the critical value. Eventually the field will move so much that it begins to feel the potential. At this point the surfing stops,  but the absolute velocity of the chameleon is large, set by the surfer solution to be $ \sim M_{\rm Pl}H/\beta$. This drives the chameleon  rapidly up the potential, so that the latter changes very quickly over time, resulting in a significant production of energetic quantum fluctuations. The classical description is lost, and with it  the ability of the theory to remain predictive as far back as BBN.

\section{DBI Correction} \label{DBIsec}

The problems caused by the surfer solution and its existence in the early Universe suggest we should seek a consistent UV modification to the chameleon model that eliminates the surfer without spoiling the interesting  low energy screening of scalar forces. In this section we will show how the DBI-chameleon model (\ref{DBI}) presented in the introduction does exactly that, for a technically natural choice of the scale of modification, $\Lambda$. Our analysis will combine both idealised dynamical systems and numerical simulations.

\subsection{Dynamical System Analysis}
Consider the modified action given by (\ref{DBI}). Introducing  $P(X)=\Lambda^{4} - \Lambda^{4}\sqrt{1-\frac{2X}{\Lambda^{4}}}$  where $X = -(\partial\phi)^2/2$, this yields the following field equations, 
\begin{equation}
3M_{\rm Pl}^2H^2 = \rho_\phi + \rho,
\label{elimination2}
\end{equation}
\begin{equation}
M_{\rm Pl}^2\left(2\dot H+3H^2 \right) = -p_\phi -p,
\end{equation}
\begin{equation}
\left(P,_{X}+2X P,_{XX}\right)\ddot{\phi} + 3HP,_{X}\dot{\phi} + V'(\phi) = -\dfrac{\beta}{M_{\rm Pl}}\rho \Sigma,
\end{equation}
where $P,_X$ denotes differentiation of $P(X)$ with respect to $X$. The energy density and pressure of the scalar  now take a more complicated form $\rho_\phi = 2XP,_X - P(X)$ and $p_\phi =P(X)$. Again we have assumed a perfect fluid with equation of state $w=p/\rho$ as our matter content, so that $\Sigma=1-3w$, and will henceforth neglect the potential $V(\phi)$. As a consistency check one can take the $\Lambda \rightarrow \infty$ limit of these equations, decoupling the DBI interactions again, and see that they give the standard chameleon equations of motion in the previous section. Eliminating $\rho$ using (\ref{elimination2}), we are left with two equations:
\begin{equation}
M_{\rm Pl}^2H^2(4-\Sigma) + 2M_{\rm Pl}^2\dot{H} + p_\phi-\rho_\phi  + \frac{(2+\Sigma)}{3}\rho_\phi =0, \showlabel{dbiH}
\end{equation}
\begin{equation}
\left(P,_X+2XP,_{XX}\right)\ddot{\phi} + 3HP,_X\dot{\phi} + 3M_{\rm Pl}\beta\Sigma H^2\left(1-\dfrac{\rho_\phi}{3M_{\rm Pl}^2H^2}\right) =0. \showlabel{dbiPhi}
\end{equation}
Introducing the same set of variables  (\ref{xdef} - \ref{zdef}) adapted to constant Jordan frame temperature at the fixed points, we arrive at the following autonomous system 
\begin{eqnarray}
\dot x &=& z \label{dbixdot},\\
\dot y &=&  \frac{1}{H_y} \left[H_z \left(3Hs^2z-\frac12 M_{\rm Pl} H^2\left[ (1-y)\left( 2+\Sigma(1-6\beta^2s^3) \right) +3y(1-s)-6(1-s^2)\right]\right) \right. \nonumber \\
&& \left.\qquad +\frac{H^2}{2} \left(\Sigma-4-y(\Sigma+2)+3y(1-s)\right) \right], ~~\\
\dot z &=& -3Hs^2z+\frac12 M_{\rm Pl} H^2\left[ (1-y)\left( 2+\Sigma(1-6\beta^2s^3) \right) +3y(1-s)-6(1-s^2)\right],\label{dbizdot}
\end{eqnarray}
where $s=\sqrt{1-\dfrac{(z-M_{\rm Pl}H)^2}{\beta^2\Lambda^4}}$. Again $H_y=\frac{\partial H}{\partial y}$, $H_z=\frac{\partial H}{\partial z}$ but now we have  $H(y, z)$ given implicitly by the following equation
\begin{equation}
\left(\Lambda^4-\dfrac{(z-M_{\rm Pl}H)^2}{\beta^2}\right)\left(1+\dfrac{3M_{\rm Pl}^2H^2y}{\Lambda^4}\right)^2 = \Lambda^4.
\label{constraintDBI}
\end{equation}
In Section \ref{chameleonsec} the surfer lay at $z=0$, with $y$ chosen  such that (\ref{constraint1}) did not constrain $H$. All other fixed points corresponded to an empty Universe with $H=0$. Such a solution would not ``surf" since there would be no cosmic expansion in Einstein frame. Therefore, to identify a generalised surfer we require $z=0$, with $y$ chosen such that (\ref{constraintDBI}) does not constrain $H$, in complete analogy to the previous case. Setting $z=0$ in (\ref{constraintDBI}), we arrive at the following polynomial in $M_{\rm Pl} H$, 
\begin{equation}
\left(\dfrac{9y^2}{\beta^2\Lambda^{8}}\right)(M_{\rm Pl}H)^6 + \left(\dfrac{6y}{\beta^2\Lambda^4}-\dfrac{9y^2}{\Lambda^4}\right)(M_{\rm Pl} H)^4 + \left(\dfrac{1}{\beta^2}-6y\right)(M_{\rm Pl} H)^2 =0.
\label{constraint2}
\end{equation}
This fails to constrain the Hubble parameter if, and only if,   all of the coefficients vanish. There are only two ways that this can happen. One possibility is to  take $\Lambda\rightarrow\infty$ and set $y=1/6\beta^2$. This is the decoupled DBI interaction limit which reduces us back to the chameleon theory, and so it's unsurprising that this leads to surfing behaviour. The other option we have is  taking $\beta\rightarrow\infty$, and setting $y=0$. This is the limit which strongly couples the scalar field to matter, with the surfer re-emerging because the kicks coming from the matter are dominant over any of the DBI self-interactions. 
In general we expect that the DBI corrections will spoil any surfing behaviour provided $\Lambda^2 \lesssim M_{\rm Pl} H/\beta$, where the scale $M_{\rm Pl}H$ is set by the scale at which the kicks occur. This follows from the fact that the DBI corrections become significant whenever $\dot \phi^2 \sim \Lambda^4$, and   on a would-be surfer, we have $\dot \phi=-M_{\rm Pl} H/\beta$.  

\begin{figure}[h!]
\centering
\begin{subfigure}{\textwidth}
  \centering
  \includegraphics[width=\linewidth]{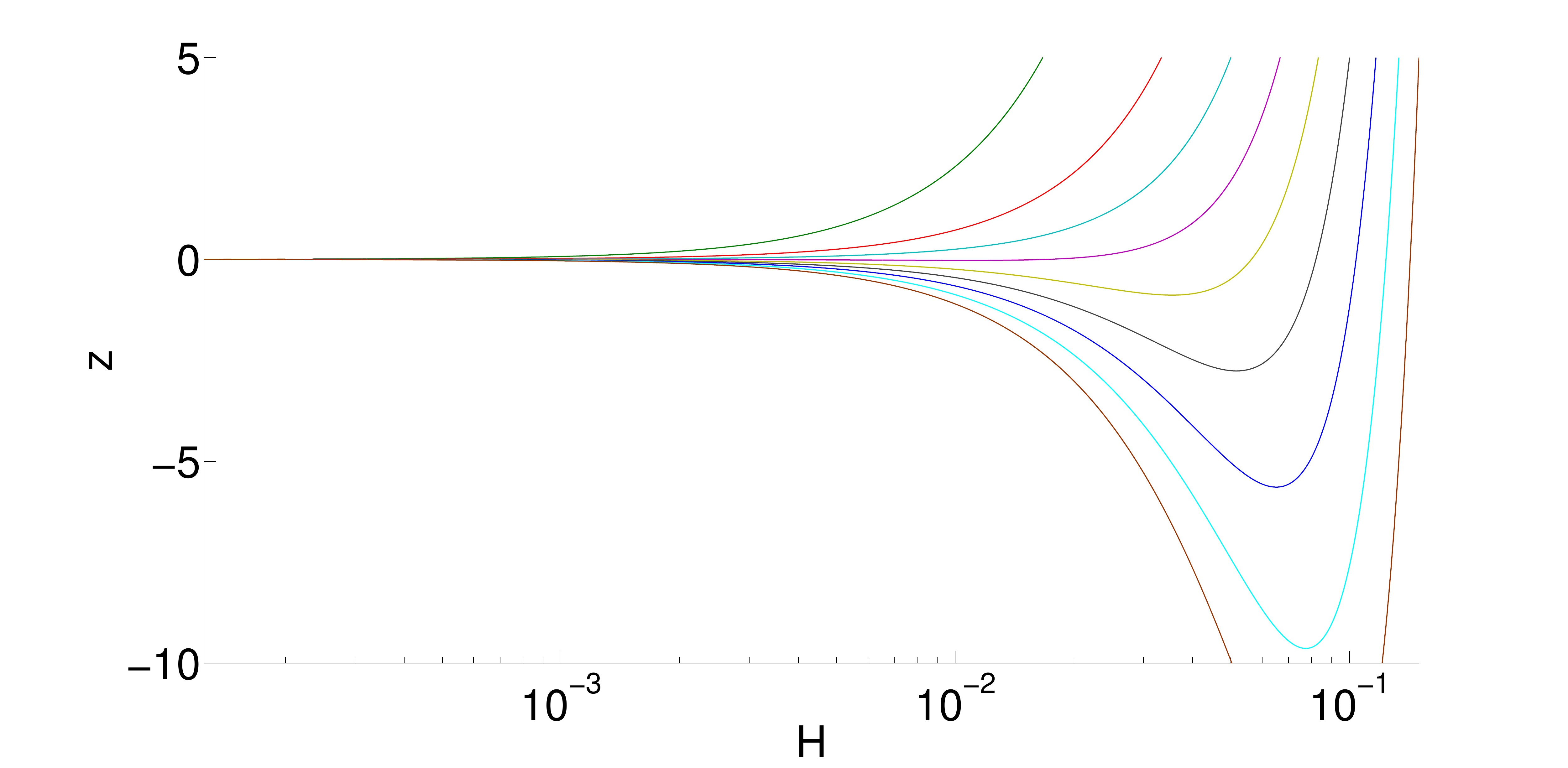}
  \caption{$\Lambda = 0.005M_p$ and $\beta = 2$.}
  \label{NormalDBI}
\end{subfigure}
\begin{subfigure}{.48\textwidth}
  \centering
  \includegraphics[width=\linewidth]{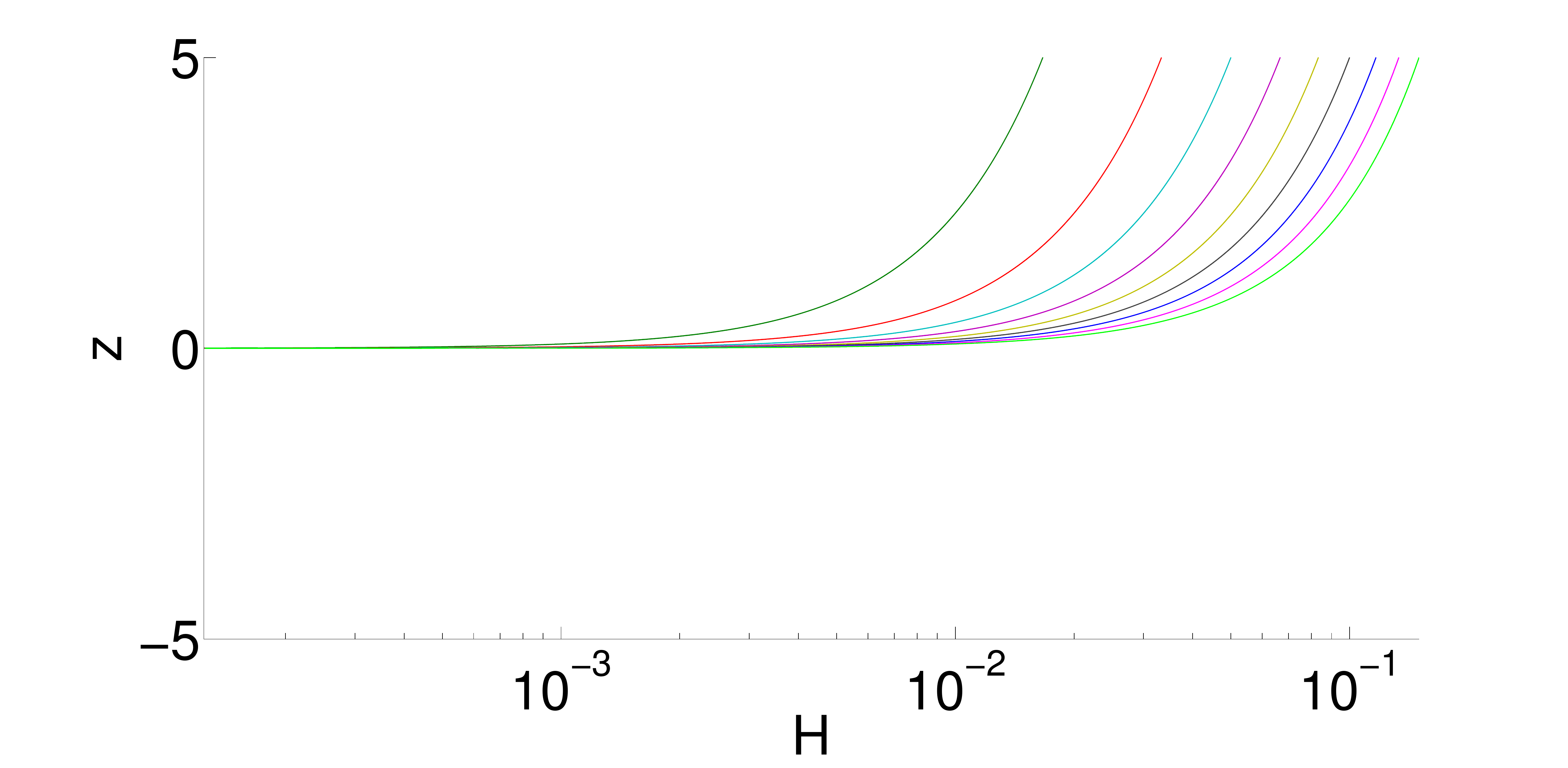}
  \caption{$\beta=20$}
  \label{DBIHighBeta}
\end{subfigure}
\begin{subfigure}{.48\textwidth}
  \centering
  \includegraphics[width=\linewidth]{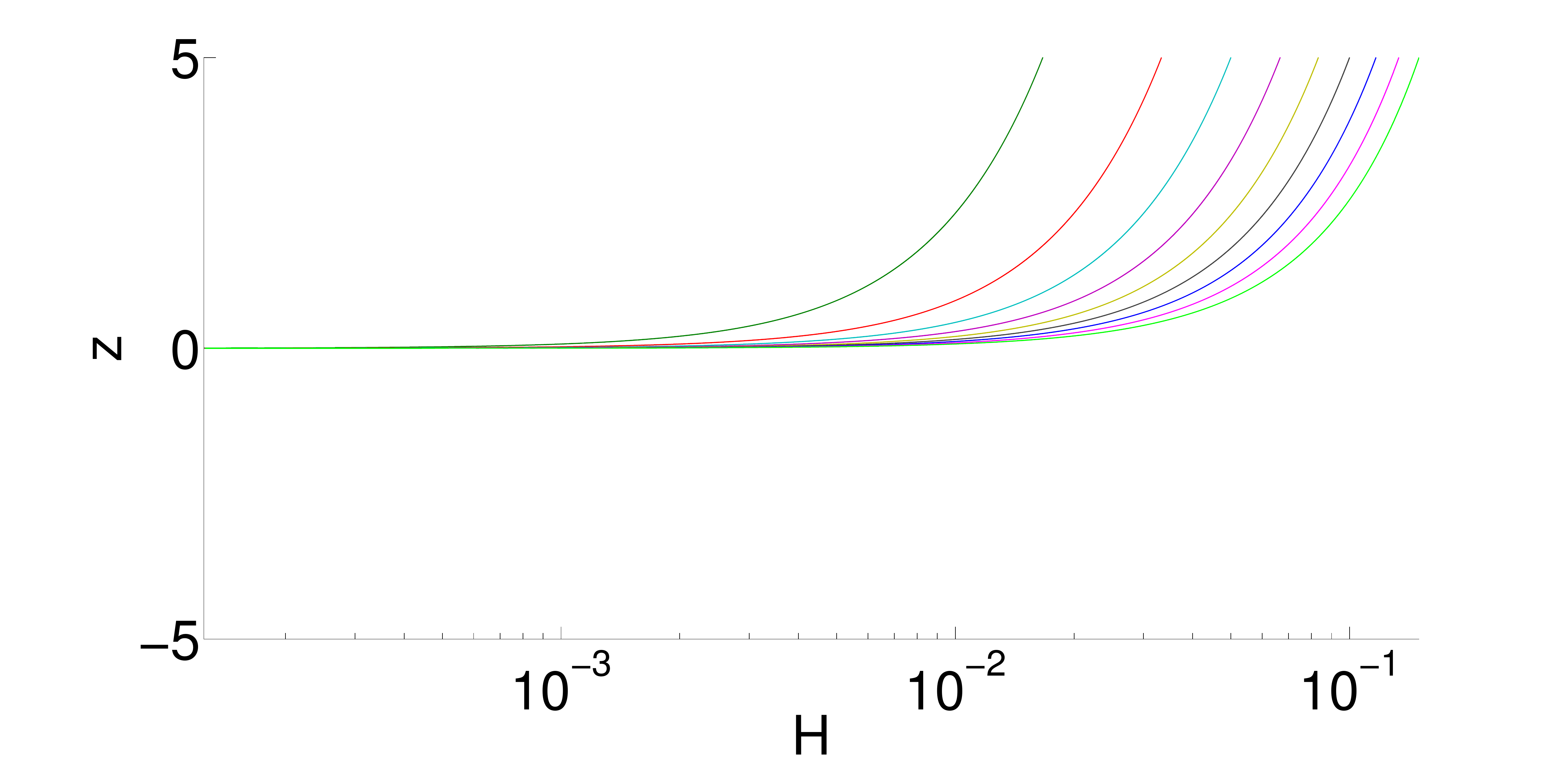}
  \caption{$\Lambda=0.5M_p$}
  \label{DBIHighLambda}
\end{subfigure}
\caption{Plots of $z$ against $H$ for a range of initial conditions, indicated by different colours. We have set $\Sigma = \dfrac{2}{6\beta^2-1}$ for all graphs, and vary $\beta$ and $\Lambda$.}
\label{DBIall}
\end{figure}

These conclusions are reinforced by solving the dynamical system numerically for a large range of initial conditions. 
When the kick function is fixed at its critical value  $\Sigma=2/(6\beta^2-1)$, we can  vary both $\beta$ and $\Lambda$ and see the surfer emerge in the limit where either $\Lambda \to \infty$ or $\beta \to \infty$.  Figure \ref{NormalDBI} shows the phase portrait for a `normal' DBI system, where $\beta$ is order one and we include $H \gtrsim  \Lambda^2/M_{\rm Pl}$. We can see that there is no longer a stable line at $z=0$ until $H$ drops below the critical value set by $\Lambda^2/M_{\rm Pl}$, which means that the surfing behaviour is no longer present in the theory at high scales. Figure \ref{DBIHighBeta} shows us how the phase portrait changes if we send $\beta$ to very large values. We can see that the stable, attractive $z=0$ line re-emerges, indicating the return of the surfer. Figure \ref{DBIHighLambda} looks very similar, in that again the surfer has returned and is stable and attractive.

The absence of the surfer in the appropriate limit of the DBI theory is expected to suppress the field excursions performed during the BBN kicks. This is crucial in order to protect the scalar from feeling the potential too soon, and at too high a velocity, resulting in significant production of high energy quanta.  We can estimate the (maximum) field excursions induced by the kicks by looking at the change in the Jordan frame temperature. Recall that the original chameleon surfer corresponded to constant Jordan frame temperature at all times, so, in the limit where the potential can be neglected, the kick will induce an infinite field excursion for suitably chosen initial conditions. In reality this is an over-estimate, but what it really tells us is that the field plunges up the potential far too early, resulting in the  breakdown of the classical EFT. To compute the corresponding field excursions for the DBI theory, we note that
\begin{equation}
\Delta\phi = \int_{t_i}^{t_i+\Delta t} \dot{\phi} \, \mathrm{d}t,
\label{deltaphi}
\end{equation}
where the kick remains ``close" to the critical value between times $t_i$ and $t_i+\Delta t$. Since the kick $\Sigma=\Sigma(T_J)$, we define ``close" in terms of the change induced in the Jordan frame temperature, 
\begin{equation}
\centering
\dfrac{\Delta T_J}{T_J} = -\int_{t_i}^{t_i + \Delta t} \dfrac{z}{M_p} \,\mathrm{d}t ,
\label{deltaT}
\end{equation}
which we will require to be $\lesssim 1$ in absolute value. The maximum field excursions will occur when the field is surfing initially, so that $z(t_i)=0$. It follows that $\dot \phi(t_i) =-M_{\rm Pl}H/\beta$, and so
\begin{equation}
\dfrac{\Delta\phi}{M_{p1} } \approx -\frac{H \Delta t}{\beta}.
\end{equation}
Using the fact that $z(t)\approx \dot z(t_i) (t-t_i)$, we further find that 
\begin{equation}
\dfrac{\Delta T_J}{T_J} \approx -\dfrac{\dot{z}(t_i) }{M_p}\dfrac{\Delta t^2}{2}.
\end{equation}
To obtain an expression for $\dot z(t_i)$, we simply set $z=0$, $y \sim {\cal O}(1)$, and $\Sigma =2/(6\beta^2-1)$ in equation (\ref{dbizdot}). This yields $\dot z(t_i) \sim {\cal O}(1) M_{\rm Pl}^3 H^4/\beta^2 \Lambda^4$, and so bringing everything together we get
\begin{equation}
\dfrac{\Delta\phi}{M_p} \approx {\cal O}(1) \frac{\Lambda^2}{M_{\rm Pl} H} \sqrt{\left|\dfrac{\Delta T_J}{T_J}\right|} \lesssim  \frac{\Lambda^2}{M_{\rm Pl} H}.
\end{equation}
Numerical simulations reinforce this estimate, although we do not present them here. It immediately shows how $\Lambda$ serves to suppress the large field excursions in comparison to the chameleon. Constraints on the variation of particle masses from BBN until today suggest that we impose an upper bound $|\Delta \phi|/M_{\rm Pl} \lesssim 0.1/\beta$ \cite{amanda}. Given that the last and most significant kick is due to the electron, at which point $M_{\rm Pl} H \sim (\text{MeV})^2$, this requires that we take $\Lambda \lesssim \text{MeV}/\sqrt{\beta}$ to protect the theory from  dangerously large field excursions.

One final thing we should check is that higher order operators invariant under the non-linear symmetry (\ref{sym}) remain suppressed, which amounts to satisfying the conditions in (\ref{higherordercorrections}).  Recall that in the limit where the DBI structure first becomes significant,  the conditions reduce to $H \ll \Lambda$. This is easily satisfied even when, in order to eliminate the surfing, we impose  $\Lambda \lesssim \sqrt{M_{\rm Pl} H/\beta}$, where $H$ is the scale of the kick. Indeed, we have explicitly  checked that  both conditions  (\ref{higherordercorrections}) hold for a range of initial conditions when we impose the ``no surfing" bound on $\Lambda$. To compliment the explicit analytic checks, we also numerically checked and found that indeed these higher order operators remain suppressed. Both of these results show us that the conditions in (\ref{higherordercorrections}) are satisfied and so the DBI effective theory described by (\ref{DBI}) remains valid.

\subsection{Numerical Simulations}
We now confirm the results of the dynamical systems analysis by performing numerical simulations, solving the system with realistic initial conditions and in the presence of a simplified kick. As was done in \cite{kick1, kick2}, we make a change of variables to ones more suited to numerical simulation by rescaling the field $\phi$ by $\mpl$ and changing the time variable to Einstein frame e-folds $N$. The equations of motion for the original chameleon theory (\er{basicH} and \er{basicPhi}) are now
\begin{align}
2 \frac{H'}{H} +  \left( 4-\Sigma \right) +  \left( \frac{2+\Sigma}{3} \right) \frac{\varphi '^2}{2} &=0,\showlabel{basicHnum}\\
\varphi'' + \left( \frac{H'}{H} +3 \right) \varphi' + 3 \beta \Sigma \left(1-\frac{\pps}{6} \right)&=0,
\showlabel{basicPhinum}
\end{align}
where in this section a prime denotes differentiation with respect to $N$.
Similarly, the equations of motion for the DBI-chameleon (\er{dbiH} and \er{dbiPhi}) become
\begin{align}
2 \frac{H'}{H}+  \left( 4-\Sigma \right) + \left[p_{\phi} - \rho_{\phi } +  \left( \frac{2+\Sigma}{3} \right) \rho_{\phi} \right] /\mpl^2H^2&=0,  \showlabel{dbiHnum}\\
\left( P,_{X} + P,_{XX} \mpl^2 H'^2 \varphi'^2 \right) \left( \varphi'' + \frac{H'}{H} \varphi' \right) + 3P,_X \varphi'  + 3 \beta \Sigma \left(1-\frac{\rho_{\phi}}{3 \mpl^2 H^2} \right)&=0. \showlabel{dbiPhinum}
\end{align} 

For simplicity, we approximate a SM kick as a single Gaussian bump in terms of Jordan frame temperature such that
\begin{align}
\Sigma(T_J) &=   A \exp{\left[-\frac{(\log T_J-  \log T_{peak})^2}{\sigma^2}\right]}. \label{kick}
\end{align}
The parameters $A$, $T_{peak}$ and $\sigma$ characterise the kick and can be appropriately chosen to model the desired kick \cite{kick2}. For the purpose of this paper we concentrate on the electron/positron kick and therefore work with $A=0.1$, $T_{peak}=2 \times 10^{-4}$ GeV and $\sigma=0.3$. Since the Jordan frame temperature goes like $T_J \sim 1/ a_J$ we can relate it to Einstein frame e-folds by
\begin{align}
T_J \left( N \right) = T_{J,i} \exp \left[ -N - \beta \Delta \varphi \left( N \right) \right] \label{Tjn},
\end{align}
where $T_{J,i}$ is the initial Jordan frame temperature corresponding to $N=0$, and $\Delta \varphi(N) = \varphi(N) - \varphi_i$ is the field excursion after $N$ Einstein frame e-folds.
Combining \er{kick} and \er{Tjn} we get an expression for the kick function in terms of Einstein frame e-folds $\Sigma(N)$, to be used in the simulations. We note that $T_J(N)$ is not necessarily decreasing. From \er{Tjn} we see that if $\varphi$ decreases faster than $N$ increases (i.e. faster than the critical velocity of the surfer), the Jordan frame will contract while the Einstein frame expands.

We solve the evolution equations \er{basicHnum} \er{basicPhinum} and \er{dbiHnum} \er{dbiPhinum} with the Gaussian kick function \er{kick}. In the DBI-chameleon runs we use various values of $\Lambda$ to demonstrate the effect of the DBI correction on the surfing solution, and to consistently compare and contrast with the chameleon we match the initial conditions for $\phi$ and $\dot{\phi}$ in each case. As long as we maintain $\gamma \sim \mathcal{O}(1)$, this ensures that the initial energy densities are comparable. In all cases we choose the matter coupling strength to be $\beta=3$ and the initial Jordan frame temperature to be $T_{J,i}=10^{-2}$GeV, approximately the temperature at which the electron/positron kick starts. We investigate positive and negative $\dot{\phi}$ and relate their energy scale to that of the kick by
\begin{align}
\dot{\phi}_i=\Lambda_k^2 \{ 0.8, -0.8\},
\end{align}
with the energy scale of the kick, $\Lambda_k = 10 T_{peak}=2 \times 10^{-3} {\rm GeV}$. When $\Lambda \approx \Lambda_k$, $\dot{\phi}$ is strongly in the DBI regime, with $\gamma \approx 1.7$ and we therefore expect the surfer to no longer exist. Neglecting any contribution to the energy density from non-relativistic matter, the initial energy density $\rho_{i}$ is dominated by radiation and is given by
\begin{align}
\rho_{r,i} = \frac{\pi^2}{30} g_{*}\left( T_{J,i} \right) T^4_{J,i} e^{4\beta \varphi_i}, \showlabel{rhori}
\end{align}
with  $g_{*}(T_{J,i})=10.75$. This closes the system of equations and allows one to calculate the three initial conditions $\varphi_i , \varphi'_i, H_i$ required to solve either \er{basicHnum} \er{basicPhinum} or \er{dbiHnum} \er{dbiPhinum}.

\begin{figure}[]
\centering
\begin{subfigure}{\textwidth}
\includegraphics[width=0.85\linewidth]{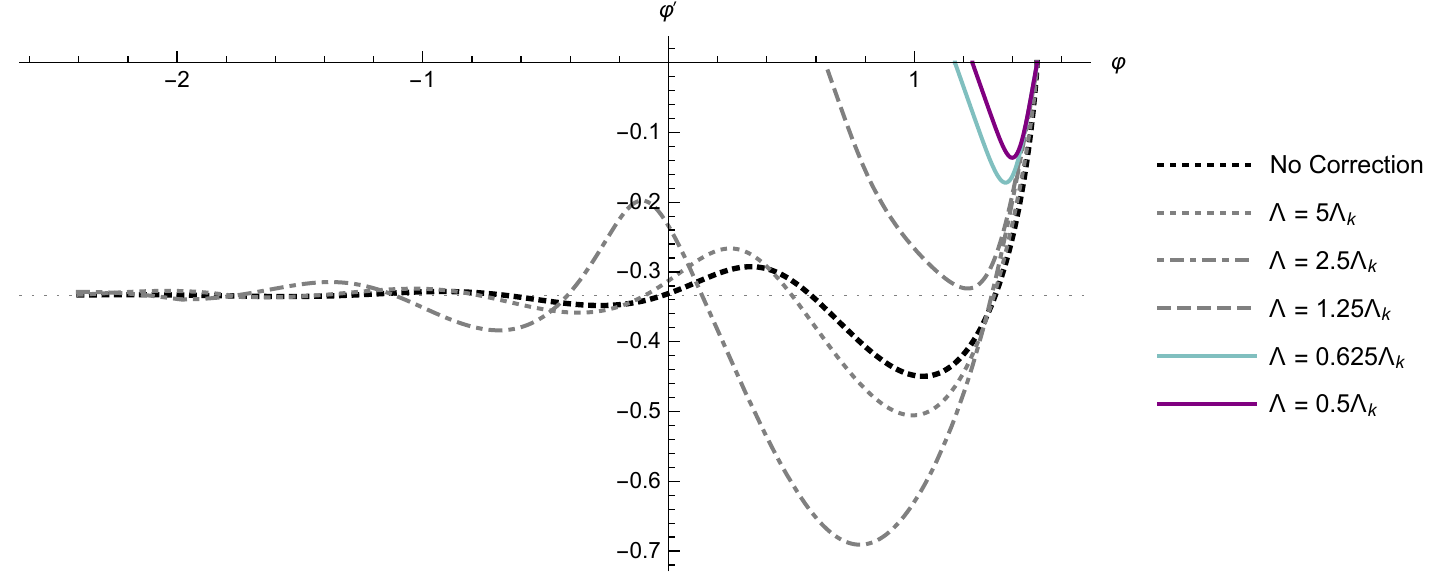}
\subcaption{Phase diagram for $\varphi'$ vs. $\varphi$. Dotted light grey line indicates the surfing solution at $-1/\beta$.}
\label{pta}
\end{subfigure}
\begin{subfigure}{\textwidth}
\includegraphics[width=0.85\linewidth]{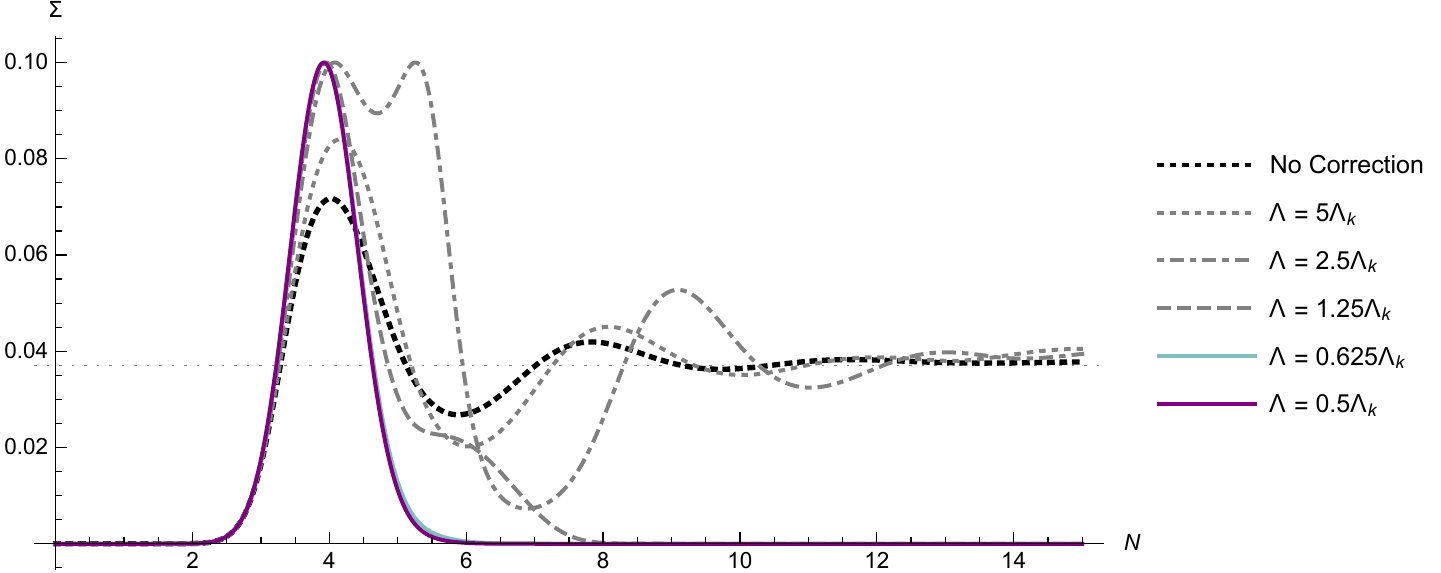}
\subcaption{The kick as a function of Einstein frame e-folds. Dotted light grey line indicates $2/(6\beta^2-1)$, the critical value of the kick at which the surf occurs. Chameleons that surf see a constant kick at this value.}
\label{ptb}
\end{subfigure}
\begin{subfigure}{\textwidth}
\includegraphics[width=0.85\linewidth]{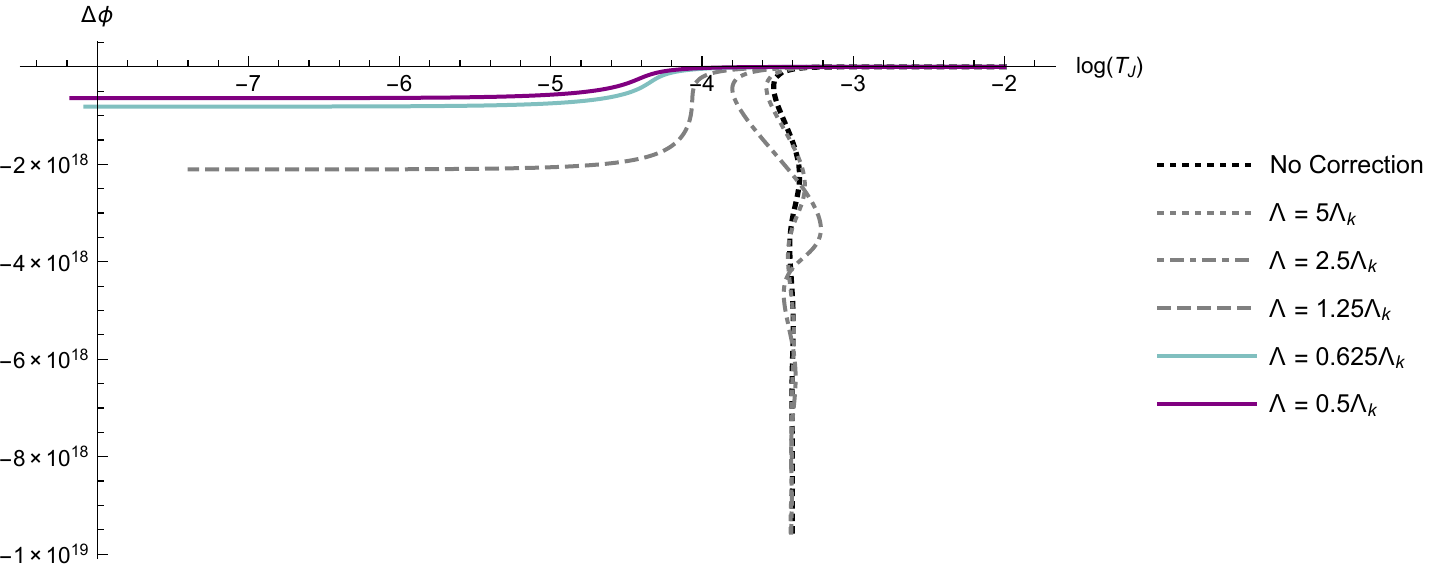}
\subcaption{Field excursion vs. Jordan frame temperature. }
\label{ptc}
\end{subfigure}
\caption{Results from the numerical situations for various values of $\Lambda$. The thick black dotted line corresponds to the original chameleon without DBI correction. Grey broken lines (dashed, dot-dashed and dotted) correspond to chameleons with DBI correction that do surf. Solid lines correspond to chameleons for which the DBI correction effectively destroys surfing behaviour.}
\label{simresults}
\end{figure}

For the DBI-chameleon we perform five runs which correspond to values of $\Lambda$ ranging from $\Lambda_{k}$ to $10 \Lambda_{k}$. The results are shown in Figure \ref{simresults}, where in each case we see that as $\Lambda \rightarrow \Lambda_k$ the surfing solution becomes unstable and eventually attractor behaviour is destroyed.  Note that for these runs, the trajectories for the positive and negative $\dot{\phi}_i$ are very similar, so we only show the positive. In all parts of Figure \ref{simresults} the thick black dotted line corresponds to the original chameleon, the grey broken lines (dashed, dot-dashed and dotted) correspond to the DBI-chameleon runs that do surf, and the coloured solid lines correspond to DBI-chameleons that don't surf. All curves represent 15 Einstein frame e-folds. 

Figure \ref{pta} shows the trajectories in the $\varphi - \varphi'$ plane with the horizontal dotted grey line indicating the surfing solution at $-1/\beta$. The effect of the DBI correction is evident in the two solid coloured curves, where field velocity decays to zero, while all other trajectories approach $-1/\beta$. Figure \ref{ptb} shows the kick as a function of Einstein frame e-folds with the horizontal dotted grey line indicating the critical value at which the surf occurs, $2/(6\beta^2-1)$. Chameleons which surf see an almost constant kick at this value, while the DBI chameleons for which the correction is effective see a kick over a finite amount of Einstein frame e-folds. 

\begin{figure}[]
\centering
\begin{subfigure}{\textwidth}
\includegraphics[width=1\linewidth]{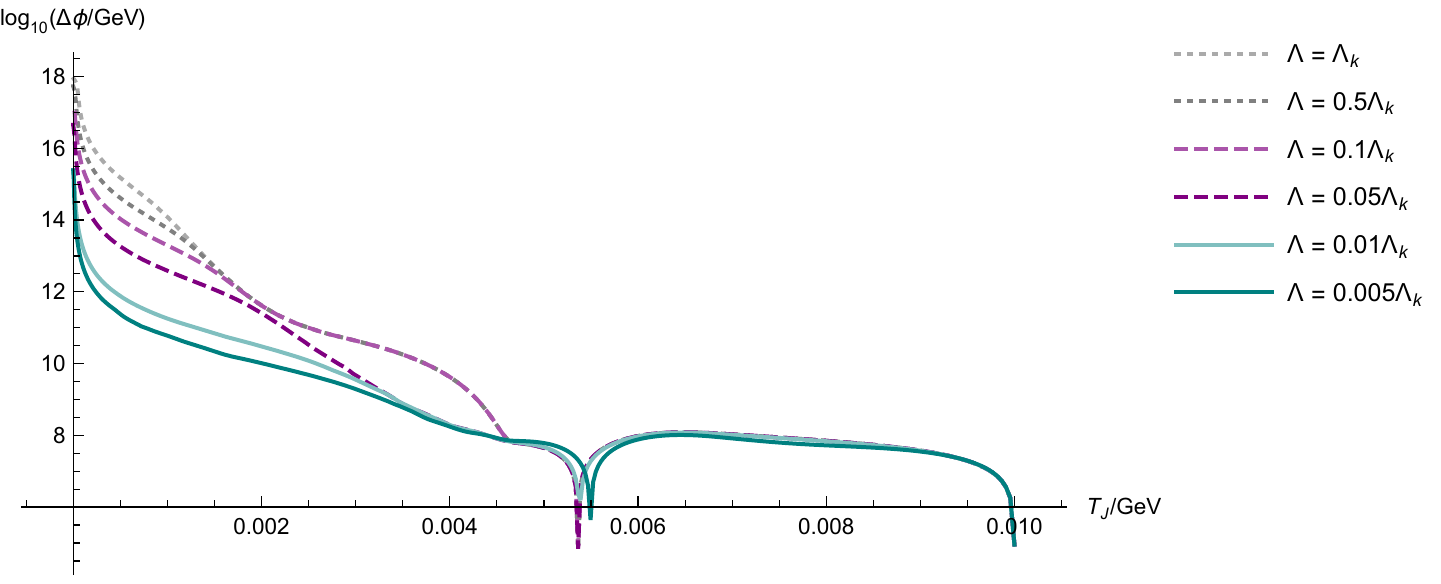}
\subcaption{Field excursions vs Jordan frame temperature, for variations in the  DBI scale at fixed initial field velocity.}
\label{fig:vain}
\end{subfigure}
\begin{subfigure}{\textwidth}
\includegraphics[width=1\linewidth]{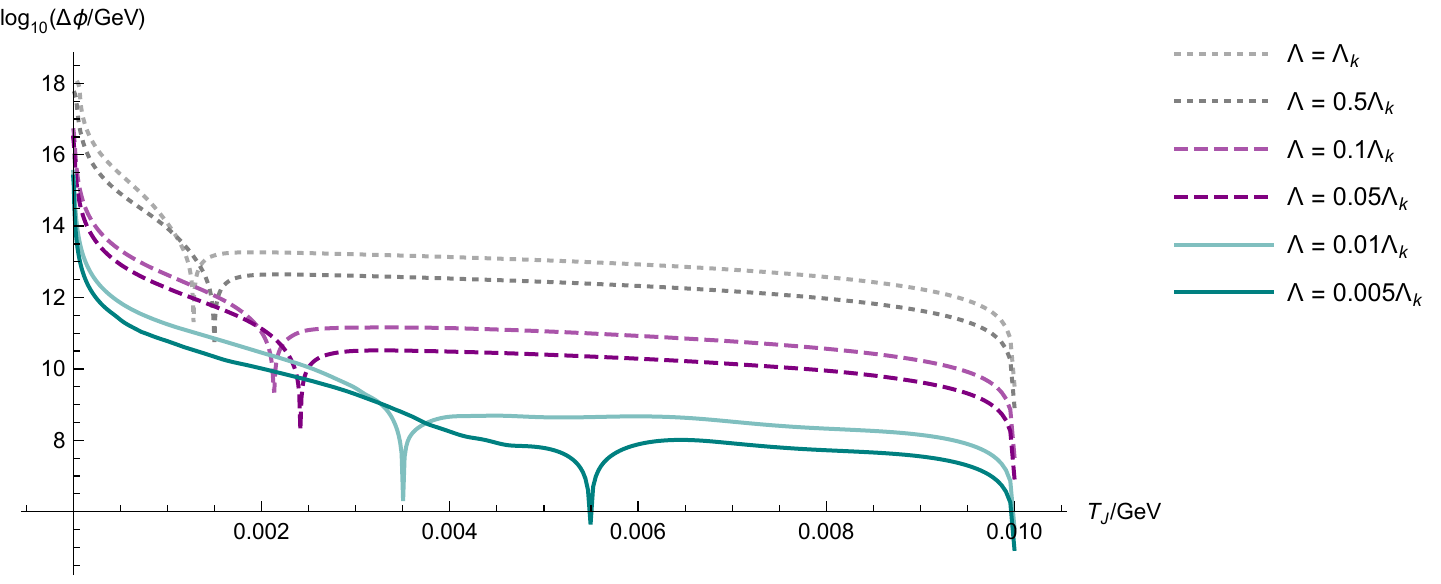}
\subcaption{Field excursions vs Jordan frame temperature for variations in the  DBI scale and the  initial field velocity, holding the effective coupling fixed.}
\label{fig:limit}
\end{subfigure}
\caption{Both plots shows the field excursions quickly becoming sub-Planckian. In the top plot this is due to a weakening of the effective coupling through a cosmological Vainshtein effect, whereas in the lower plot this is due to the DBI structure imposing a `cosmic speed limit' on the scalar. }
\label{ }
\end{figure}

Figure \ref{ptc} shows the field excursion against Jordan frame temperature, with the vertical line indicating the critical temperature of the surf, $T^c_J$. The solid coloured lines show the effect on the DBI correction on the field excursion, with stronger corrections reducing the excursion in the field.  Although it is clear that the surfing behaviour is ultimately eliminated by the DBI correction,  we might still worry that Planckian field excursions are still possible. This is problematic if we impose the BBN bound that $\phi <0.1 M_{pl}/\beta$ just before the electron kick, since it would mean the scalar still crashed into the minimum of the effective potential during BBN, with too large a velocity. However, it is important to realise that these plots represent a worst case scenario, with the DBI scale close to the kick scale, and the initial speed of the chameleon close to the speed limit.  We can easily suppress the field excursions to safe values  by lowering the DBI scale by just an order of magnitude further. This is demonstrated in figures \ref{fig:vain} and \ref{fig:limit}.  In figure  \ref{fig:vain}, the initial field velocity is held fixed for each run, whilst the DBI scale is lowered. This means the $\gamma$ factor is increasing, and so the effective coupling is being weakened, suppressing the excursions generated by the kick.  In figure  \ref{fig:limit} we lower the DBI scale and the initial velocity in tandem, such that the $\gamma$ factor  is held fixed.  This time the effective coupling is unchanged between runs, but the field excursions are suppressed thanks to the decreasing initial velocity. This reflects the ability of the DBI structure to impose a cosmic speed limit on the scalar field. Therefore, even if we choose the initial value of $\phi <0.1 M_{pl}/\beta$ just before the electron kick, in accordance with BBN bounds on the variation of particle masses, we are no longer doomed to crash into the minimum of the effective potential during BBN with too large a velocity, thanks to the DBI suppression of field excursions. Note that in all of our simulations the higher order corrections (\ref{higherordercorrections}) remain suppressed.

\newpage
\section{Conclusion} \label{conclusionsec}

It has recently been claimed that for some set of initial conditions the chameleon is not a consistent classical field theory for describing early universe cosmology. During the radiation dominated era significant variations in $\Sigma = (\rho - 3P)/ \rho$ due to massive particles becoming non relativistic have drastic implications for the chameleons evolution and the production of highly energetic quantum fluctuations invalidates any classical treatment of the chameleon. This suggests that the chameleon model cannot remain predictive as far back as BBN unless we demand that the field is sufficiently weakly coupled to matter. One might argue that although perturbation theory breaks down at the bounce, the bounce process itself is perfectly classical and so indeed there is no issue. To some extent this is an unknown.   We do not have access to repeatable empirical data explicitly confirming the reliability of the classical theory through the bounce, therefore we require knowledge of the theory beyond the perturbative regime, asking questions of an as yet unknown UV completion.

In any case, to remove any element of doubt we have shown that considering a chameleon theory as an EFT to be corrected by a DBI term at high velocities protects the field from dangerously large field excursions. The DBI correction ensures that the chameleon's effective coupling to matter is weakened {\it dynamically} in the early Universe, such that the impact of SM kicks are suppressed. The mechanism by which this is achieved is similar to the Vainshtein mechanism, and exploits derivative self interactions of the scalar in order to induce a large $Z$ factor on non-trivial homogeneous solutions.  Working with a DBI correction to the chameleon, we have re-analysed the scalar evolution  in the presence of  kicks. For suitably chosen (but technically natural) parameters 
\begin{equation} \label{bound}
\frac{(\text{MeV})^2}{M_{\rm Pl}} \ll \Lambda \lesssim \frac{\text{MeV}}{\sqrt{\beta}}
\end{equation}
we found that the non-linear derivative interactions dominate the dynamics as SM particles become non-relativistic, such that  the scalar is no longer able to surf the kicks through a large field range. Furthermore, by pushing  the DBI to just an order of magnitude below the scale of the final kick we are able keep field excursions sub-Planckian (up to factors of $\beta$), so that one can easily avoid crashing into the minimum of the effective potential during BBN. This ensures that the DBI chameleon will generically avoid  the breakdown described in \cite{kick1,kick2} for a suitably chosen DBI scale. We do not expect that this behaviour is unique to the particular UV correction we have studied here but at least within the k-essence models it is the only correction that does the job in a controllable manner, whereby higher order operators can be consistently neglected. This is the origin of the lower bound in (\ref{bound}). The presence of this lower bound also translates into a bound on $\beta \ll 10^{21}$, but it is inconceivably mild (!), in contrast to the bound imposed in \cite{kick1,kick2} or indeed the experimental bounds which are currently set at $\beta \lesssim 10^{9}$ \cite{Jenke}. 

In summary, the DBI chameleon retains the neat behaviour of the chameleon at low energies while rendering early universe standard model kicks harmless. It is important to realise that the long range scalar force coming from (\ref{DBI}) is prevented from affecting local gravitational experiments thanks to the original chameleon mechanism. The derivative interactions play no role in this regard since they are completely subdominant in the infrared. They become important at higher energies, in the early Universe, exploiting a homogeneous version of the Vainshtein mechanism to suppress the effective coupling between the scalar and non-relativistic matter. This weakens the adverse effects of the kicks, allowing us to trust the theory in the early universe without having to fine tune the strength of the scalar-matter coupling.  This suggests that the DBI chameleon is better placed than the original chameleon to describe physics throughout our cosmic history.

\section{Acknowledgements}
AP is funded by a Royal Society University Research Fellowship. DS and TW are funded by STFC studentships. This work is based on the research supported by the South African Research Chairs Initiative of the Department of Science and Technology and National Research Foundation of South Africa as well as the Competitive Programme for Rated Researchers (Grant Number 91552) (A Weltman and EP) and Innovation PhD studentship (A Walters). EP is also supported by a Masters Bursary from the South African National Institute for Theoretical Physics (NITheP).  Any opinion, finding and conclusion or recommendation expressed in this material is that of the authors and the NRF does not accept any liability in this regard. We would also like to thank C Burrage, A Erickcek, N Kaloper, J Khoury, J Shock and especially V Sivanesan for helpful discussions on this work.

\end{document}